\documentclass[%
 prx,
 amsmath,amssymb,
 preprint,
 endfloats*,
 superscriptaddress,
 nofootinbib,
]{revtex4-1}

\usepackage{float}
\usepackage{textcomp}
\usepackage{lipsum}
\usepackage{graphicx}
\usepackage{dcolumn}
\usepackage{bm}
\usepackage{textgreek}
\usepackage[utf8]{inputenc}
\usepackage[T1]{fontenc}
\usepackage{mathptmx}
\usepackage{newtxmath}
\renewcommand*{\thefootnote}{\alph{footnote}}
\begin{document}

\title{Molecular Beam Epitaxy Growth of Antiferromagnetic Kagome Metal FeSn}

\author{Hisashi Inoue$^{\rm 1,a,b}$, Minyong Han$^{\rm 1,a}$, Linda Ye$^1$, Takehito Suzuki$^1$, and Joseph. G. Checkelsky$^{\rm 1,c}$}
\noaffiliation
\affiliation{Department of Physics, Massachusetts Institute of Technology, Cambridge, Massachusetts 02139, USA}

\date{\today}
\footnotetext[1]{These authors contributed equally to the work.}
\footnotetext[2]{Present address: Frontier Research Institute for Interdisciplinary Sciences and Institute for Materials Research, Tohoku University, Miyagi 980-8577, Japan}
\footnotetext[3]{Electronic mail: checkelsky@mit.edu}

\begin{abstract}
FeSn is a room-temperature antiferromagnet expected to host Dirac fermions in its electronic structure. The interplay of magnetic degree of freedom and the Dirac fermions makes FeSn an attractive platform for spintronics and electronic devices. While stabilization of thin film FeSn is needed for the development of such devices, there exist no previous report of epitaxial growth of single crystalline FeSn. Here we report the realization of epitaxial thin films of FeSn (001) grown by molecular beam epitaxy on single crystal SrTiO$_{3}$ (111) substrates. By combining X-ray diffraction, electrical transport, and torque magnetometry measurements, we demonstrate the high quality of these films with the residual resistivity ratio $\rho_{xx}(300 \hspace{0.2em}{\rm K})/\rho_{xx}(2 \hspace{0.2em}{\rm K}) = 24$ and antiferromagnetic ordering at $T_{\rm N} = 353$ K. These developments open a pathway to manipulate the Dirac fermions in FeSn by both magnetic interactions and the electronic field effect for use in antiferromagnetic spintronics devices.
\end{abstract}

\maketitle

\begin{figure}
\includegraphics[width=16cm]{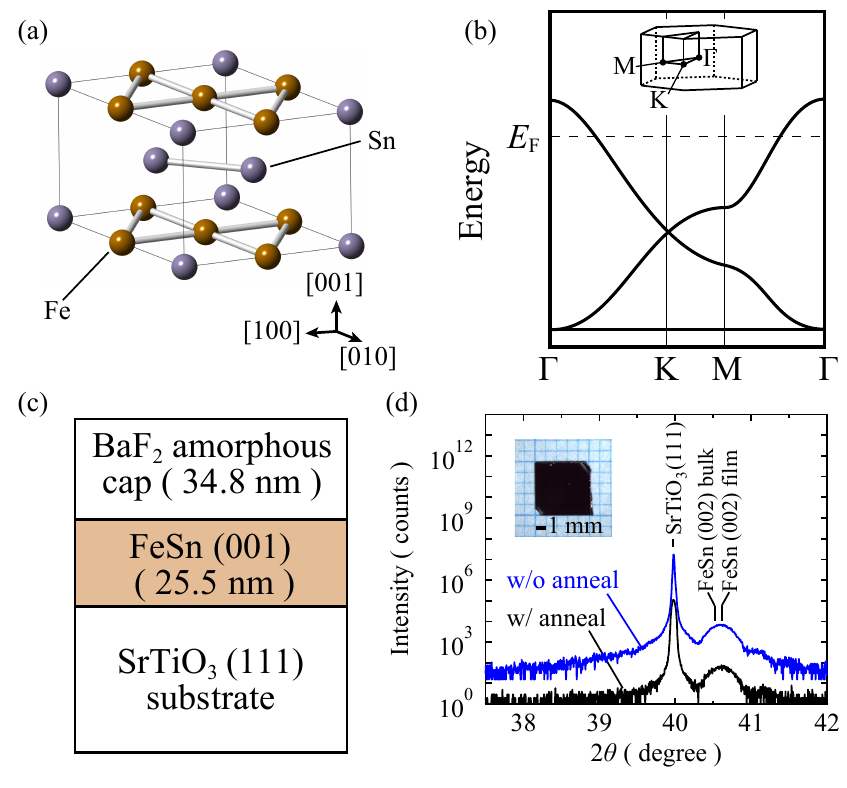}
\caption{\label{fig1} 
(Color online) (a) Crystal structure of FeSn. (b) Schematic electronic band structure of Fe kagome network. The inset shows the Brillouin zone. (c) Schematic of the thin film sample structure. (d) X-ray diffraction spectra of FeSn thin films with and without post-annealing. A vertical offset is added for clarity. Inset: Optical micrograph of an FeSn thin film with the BaF$_2$ cap and post-annealing (scale bar: 1 mm).
}
\end{figure}

The antiferromagnetic metal FeSn consists of two-dimensional layers of corner-sharing triangle network of Fe, separated by honeycomb lattices of Sn [Fig$.$ \ref{fig1} (a)]. This geometrical configuration of Fe, called the kagome lattice, is expected to host a linearly dispersing Dirac band and a topological flat band in its electronic band structure [Fig$.$ \ref{fig1} (b)] \cite{Ye2018a,Ye2018,Yin2018,Yin2019}. As the Dirac band and the flat band have been the platform for a number of intriguing physical phenomena arising from electronic correlation and the band topology \cite{Tang2011,Mazin2014}, FeSn is a material platform in which the interplay between magnetism and topology can be explored.

To study the physics of the magnetic kagome lattice in FeSn and for its electronics applications, it is desirable to realize the material in a thin film form so that it can be processed into device structures and its physical properties can be tuned electrostatically. Here, we report the first realization of high quality magnetic FeSn thin films grown by molecular beam epitaxy. X-ray diffraction measurements indicate formation of single crystalline FeSn with sharp interfaces. Our capping and post-annealing procedures result in improved quality of the films as indicated by metallic electrical transport with residual resistivity ratio of 24. Furthermore, torque magnetometry measurements of these films confirm long-range antiferromagnetic order almost unchanged from that of bulk FeSn single crystals.

FeSn thin films were grown on single crystal SrTiO$_3$ (111) substrates (Shinkosha, Co.) [Fig$.$ \ref{fig1} (c)]. Before being loaded into the growth chamber, the substrates were  cleaned with acetone and methanol, and then annealed at 1050 $^\circ$C in air for 1 hour, followed by sonication in pure water for 30 seconds at room temperature. We repeated the annealing and sonication procedures twice in order to prepare a flat surface suitable for epitaxial film growth \cite{Connell2012,Hallsteinsen2016,Woo2015}. After loading to the growth chamber, we pre-annealed the substrates at 600 $^\circ$C for 1 hour to remove any residual moisture and adsorbates. FeSn was deposited for 40 minutes by thermally evaporating Fe and Sn from solid sources using effusion cells. The substrate temperature during deposition was 150 $^\circ$C. The ratio of beam-equivalent pressures (BEPs) was $P_{\rm Fe} : P_{\rm Sn} = 1 : 2.2$, where $P_{\rm Fe}$ and $P_{\rm Sn}$ are BEPs for Fe and Sn, respectively. After the deposition, some films were capped with amorphous BaF$_2$, deposited at 200 $^\circ$C for 30 minutes. Finally, these films were post-annealed at 500 $^\circ$C for 12 hours to improve crystalline quality.

Figure \ref{fig1} (d) shows X-ray diffraction spectra of samples with and without the BaF$_2$ cap and post-annealing, where the wavelength of the incident X-ray beam was $\lambda = 0.154$ nm. They show a film peak at $2\theta = 40.61^\circ$ for the annealed sample and $2\theta = 40.60^\circ$ for the unannealed sample. These are close to the FeSn (002) peak position $2\theta = 40.52^\circ$ expected for a bulk single crystal ($c_{\rm bulk} = 0.445$ nm), confirming the formation of epitaxial single crystalline FeSn. The shift of the peak position from that of the bulk single crystal reflects the residual epitaxial strain of 0.2$\%$ from the substrate. The film peak accompanies Laue interference fringes, indicating sharp interfaces. For the scattering geometry with the scattering vector perpendicular to the sample plane, we did not observe peaks other than SrTiO$_3$ $(lll)$ and FeSn $(00l)$, where $l$ is an integer.

\begin{figure}
\includegraphics[width=16cm]{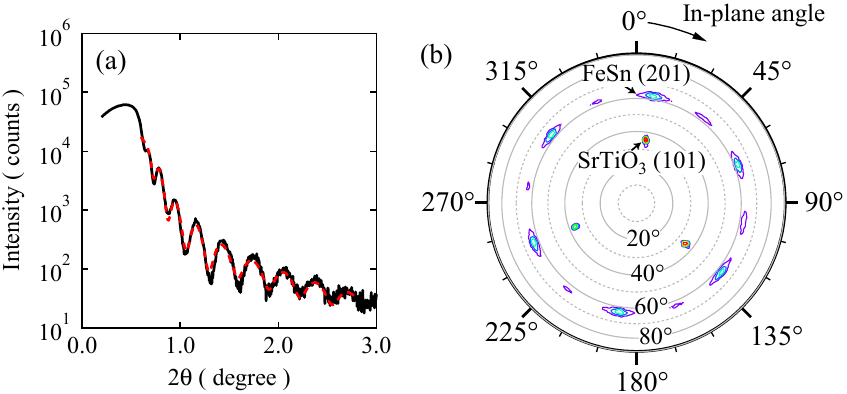}
\caption{\label{fig2}
(Color online) (a) X-ray reflectivity oscillations measured on FeSn with BaF$_2$ cap and the post-annealing process. The best fit to the data is shown as a dashed curve. (b) Pole figure plotted as a contour plot in a log scale, showing the in-plane orientation of the SrTiO$_3$ substrate and FeSn film. Numbers on the radial axis are the inclination angles of the diffraction plane normal with respect to the sample normal.
}
\end{figure}

In order to estimate the film thickness, we performed X-ray reflectivity measurements on a capped and annealed sample (see Fig$.$ \ref{fig2} (a)). The spectrum shows clear oscillations due to the interference of reflected X-ray beams indicating a flat film. By comparing the reflectivity data to a simulated relectivity curve using the model structure shown in Fig$.$ \ref{fig1} (c), we determined the thickness of FeSn and BaF$_2$ cap to be 25.5 nm and 34.8 nm, respectively. We use these estimates for thicknesses hereafter.

The in-plane orientation of the FeSn thin film with respect to the SrTiO$_3$ (111) substrate was determined from measurements of the FeSn \{201\} peaks and SrTiO$_3$ \{101\} peaks, shown as a pole figure in Fig$.$ \ref{fig2} (b). The FeSn \{201\} peaks exhibit six-fold rotation symmetry while the SrTiO$_3$ \{101\} peaks show three-fold rotation symmetry as expected from their crystal structures. The in-plane angle of the FeSn (201) peak matches with the angle of SrTiO$_3$ (101), indicating that the in-plane crystal axes of FeSn and SrTiO$_3$ are aligned at the FeSn$-$SrTiO$_3$ interface. We observe a small but finite response between the FeSn \{201\} peaks. We attribute this to formation of a minor crystal domain which is rotated by 30$^\circ$ in the in-plane direction.

\begin{figure}
\includegraphics[width=16cm]{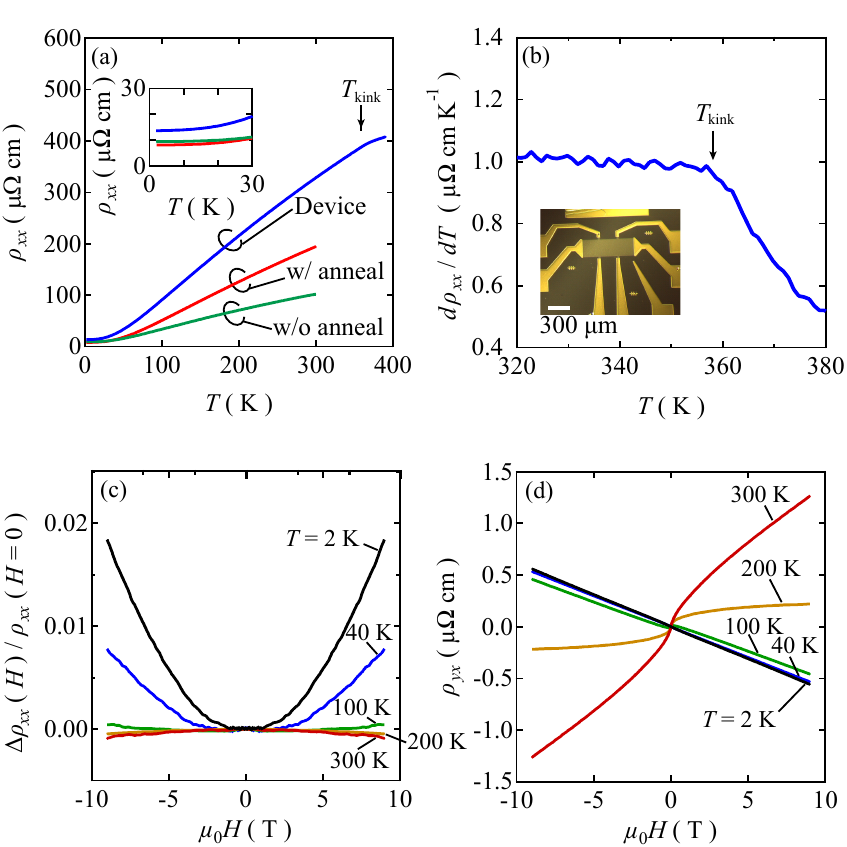}
\caption{\label{fig3}
(Color online) (a) Temperature dependence of resistivity $\rho_{xx} (T)$ of the bare FeSn film sample with post-annealing, the bare film sample without post-annealing, and the Hall-bar device sample. Inset: Magnified view of $\rho_{xx} (T)$ at low temperature. (b) Derivative of $\rho_{xx} (T)$ with respect to $T$ of the Hall-bar device. Inset: optical micrograph of the Hall-bar device (scale bar: 300 {\textmu}m). (c) Magneto-resistance and (d) Hall effect of the sample with post-annealing at selected temperatures. $\mu_0$ is the vacuum permeability, $H$ is the magnetic field, and $\Delta \rho_{xx} = \rho_{xx}(H) - \rho_{xx}(0)$. 
}
\end{figure}

For a reliable characterization of transport properties, the films were processed into Hall-bar devices. An optical microscope image of a device is shown in Fig$.$ \ref{fig3} (b) inset. The film was first patterned into a Hall-bar shape with photolithography followed by Ar milling. The milling was stopped at the FeSn / SrTiO$_3$ interface by using a precisely calibrated milling rate (to prevent damage to the substrate). In the second step, edge contacts to the FeSn film were made by depositing Ti / Au using electron beam evaporation at an angle 15$^\circ$ away from the sample normal direction. The thicknesses for Ti and Au were 7 nm and 70 nm, respectively. Subsequent electrical contacts were made by Ag paint. The contact resistance was approximately 2 $\Omega$ at temperature $T=2$ K.

Figure \ref{fig3} (a) shows the temperature dependence of the resistivity $\rho_{xx} (T)$ of three different FeSn thin film samples: a Hall-bar device, a rectangular-shaped bare film with the BaF$_2$ cap and post-annealing process, and a bare film without the BaF$_2$ cap or the post-annealing process. The thickness of FeSn layer in all these samples was 25.5 nm. All samples showed metallic behavior with $\rho_{xx}$ monotonically decreasing as temperature decreases. The resistivities at 300 K (2 K) of the bare films with and without the post-annealing process were 194 {\textmu}$\Omega$ cm (8.1 {\textmu}$\Omega$ cm) and 102 {\textmu}$\Omega$ cm (9.5 {\textmu}$\Omega$ cm), respectively. This gives residual resistivity ratio, $RRR=\rho_{xx}(300 \hspace{0.2em}{\rm K})/\rho_{xx}(2 \hspace{0.2em}{\rm K})$, of $RRR=24$ for the film with post-annealing and $RRR=10.7$ for the film without post-annealing. The factor of 2 increase in $RRR$ signifies the improved quality of the FeSn films after the post-annealing process. The resistivity of the Hall-bar device at 300 K (2 K) was 328 {\textmu}$\Omega$ cm (13.7 {\textmu}$\Omega$ cm). The Hall-bar device exhibits $RRR=24$, identical to that of the bare film with post-annealing. This indicates that the quality of the sample was unaffected by the device fabrication procedures.

A close inspection of $\rho_{xx} (T)$ of the Hall-bar device reveals a kink in the curve around $T=358$ K. To illustrate this more clearly, the derivative of $\rho_{xx} (T)$ as a function of temperature is shown in Fig$.$ \ref{fig3} (b). $d\rho_{xx}/dT$ shows a clear feature at $T_{\rm kink}=358$ K. A similar behavior of $\rho_{xx} (T)$ has been reported for FeSn bulk single crystals and associated with an onset of the antiferromagnetic transition \cite{Stenstrom1972}. The correlation of this behavior with the magnetic phase transition in our FeSn film is discussed below.

Figure \ref{fig3} (c) shows the magneto-resistance of the post-annealed sample. Magnetic fields were applied perpendicular to the sample plane. At room temperature, we see a small quadratic negative magneto-resistance, which is suppressed as temperature decreases and becomes positive below $T = 100$ K. As we will show below, our FeSn thin films exhibit antiferromagnetic order at room temperature. Therefore it is likely that the quadratic negative magneto-resistance of our FeSn thin film arises due to modulation of resistance by the antiferromagnetic order, while the low-temperature positive magneto-resistance is induced by the Lorentz force \cite{Usami1978}.

The Hall curves of the sample with post-annealing exhibit a characteristic change of the sign of slopes as temperature decreases (see Fig$.$ \ref{fig3} (d)). The high field slope $d\rho_{yx}/dH$ changes the sign from positive to negative around $T = 200$ K, and the low field slope changes the sign from positive to negative around $T = 60$ K. If we assume that only one band is occupied, this would indicate a carrier density change from 5.7 $\times 10^{21}$ cm$^{-3}$ (holes) to 9.9 $\times 10^{21}$ cm$^{-3}$ (electrons) from the high field Hall slope. Such a large carrier density change with temperature is unlikely; since Hall curves in Fig$.$ \ref{fig3} (d) shows clear non-linearity as a function of magnetic field, we attribute the Hall slope change to multi-band transport. This multi-band nature likely arises from the three-dimensional network of Sn in this material \cite{Ye2018a,Ye2018}. 

\begin{figure}
\includegraphics[width=16cm]{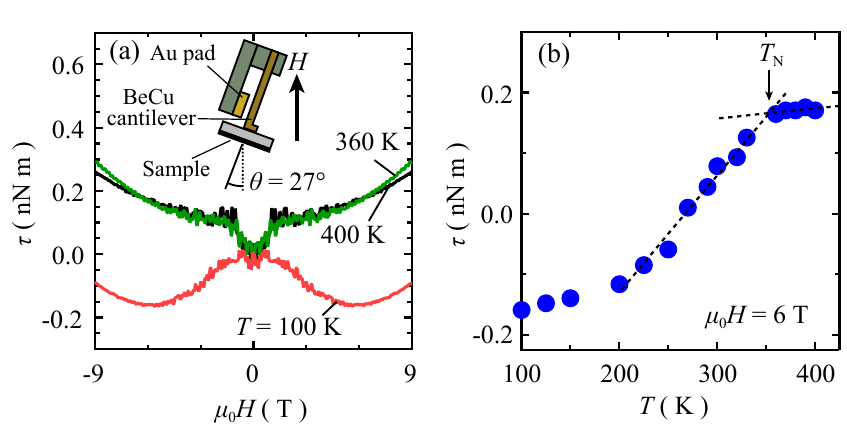}
\caption{\label{fig4}
(Color online) (a) Magnetic torque $\tau$ as a function of magnetic field. Inset: Schematic of the measurement setup. (b) Temperature dependence of $\tau$ at $\mu_0 H = 6$ T. Dashed lines are linear fits to the data for $T \leq 330$ K and $T \geq 360$ K, whose intersection gives an estimate for $T_{\rm N}$. 
}
\end{figure}

Bulk single crystals of FeSn are known to host antiferromagnetism below 368 K \cite{Hartmann1987,Yamamoto1966,Djega-Mariadassou1966}. The moments are ferromagnetically aligned within the (001) plane and antiferromagnetically stacked along the [001] direction in the antiferromagnetic phase \cite{Yamaguchi1967}. The spin direction is found to lie within the (001) plane \cite{Kulshreshtha1981,Haggstrom1975}. To confirm that antiferromagnetism appears in our FeSn thin films, we performed capacitive torque magnetometry measurements. A schematic of the measurement setup is shown in the inset of Fig$.$ \ref{fig4} (a). A FeSn thin film sample was attached to a 10 {\textmu}m-thick BeCu cantilever and a magnetic field was applied at an angle $\theta \approx 27^{\circ}$ from the sample normal. A magnetic torque ${\bm \tau} = V {\bm M} \times {\bm B}$ is generated, and the consequent deflection of the cantilever was probed by the change in the capacitance $\Delta C$ between the cantilever and a fixed Au pad, where $V$ is the sample volume, ${\bm M}$ is the magnetization, and ${\bm B}$ is the magnetic flux density. $\Delta C$ was converted to $\tau$ using the geometry and the Young's modulus of the BeCu cantilever. In the absence of any magnetic anisotropy, $\tau = 0$ because ${\bm M}$ aligns with ${\bm B}$, and $\tau$ is sensitive to the magnetic anisotropy of the sample.

In Fig$.$ \ref{fig4} (a), we plot $\tau(H)$ of a 25.5 nm-thick FeSn film with the BaF$_2$ cap and post-annealing. Above \textit{T} = 360 K, $\tau(H)$ exhibits a quadratic response with nearly temperature independent positive curvature $\tau(H) \propto H^{2}$ for $\mu_0 H > 2$ T. Such a response is characteristic of a paramagnet \cite{Wang2005}. On the other hand, below \textit{T} = 360 K, $\tau(H)$ starts to deviate from a simple parabola and at 100 K it develops a negative dip around $\mu_0 H = 6$ T. We note that similar W-shaped torque response was also observed in thin films of antiferromagnetic GdBi below the N\'eel temperature \cite{Inoue2019}. We attribute the cusp feature of $\tau(H)$ for $|\mu_0 H| < 2$ T to a mechanical instability of the BeCu cantilever.

In Fig$.$ \ref{fig4} (b), we show temperature dependence of $\tau$ at 6 T. At 360 K, $\tau(T)$ shows a kink, suggesting that an additional magnetic anisotropy developed below $T = 360$ K. We attribute this feature to the appearance of an antiferromagnetic order in the FeSn film. By linearly extrapolating $\tau(T)$ below $T \leq 330$ K and above $T \geq 360$ K, the N\'eel temperature of the film is given as the intersection of these lines $T_{\rm N} = 353$ K, which is close to the N\'eel temperature 368 K reported for FeSn bulk single crystals \cite{Hartmann1987,Yamamoto1966,Djega-Mariadassou1966}.

Finally we comment on the kink feature observed in $d\rho_{xx}/dT$ in Fig$.$ \ref{fig3} (b). The temperature at which the kink occurs $T_{\rm kink} = 358$ K is close to the temperature of magnetic transition $T_{\rm N} = 353$ K determined from the magnetic torque measurements. This suggests that the antiferromagnetic ordering of FeSn thin film gives rise to the feature in $\rho_{xx} (T)$ around $T_{\rm kink}$.

In conclusion, we report the first successful growth and characterization of epitaxial thin films of FeSn, an antiferromagnetic kagome metal. By employing controlled growth by molecular beam epitaxy, and a cap-and-post-annaeling procedure, we established a method to fabricate high quality FeSn thin films with $RRR = 24$ as confirmed by X-ray and electrical transport measurements. Stable antiferromagnetic order in our thin film at room temperature provides an opportunity to control the Dirac electronic properties by its magnetism as well as field-effect-gating for electronic and spintronics applications \cite{Smejkal2017,Smejkal2018}. 

\begin{acknowledgments}
We are grateful to R. Comin, M. Kang, J. van den Brink, S. Fang, and M. P. Ghimire for fruitful discussions. This research was funded, in part, by the Gordon and Betty Moore Foundation EPiQS Initiative, Grant No. GBMF3848 to J.G.C. and ARO Grant No. W911NF-16-1-0034. L.Y. acknowledges support by the STC Center for Integrated Quantum Materials, NSF grant number DMR-1231319, and the Tsinghua Education Foundation. The authors acknowledge characterization facility support provided by the Materials Research Laboratory at Massachusetts Institute of Technology, as well as fabrication facility support by the Microsystems Technology Laboratories at Massachusetts Institute of Technology.
\end{acknowledgments}

\bibliography{MBE_FeSn}

\begin{thebibliography}{21}%
\makeatletter
\providecommand \@ifxundefined [1]{%
 \@ifx{#1\undefined}
}%
\providecommand \@ifnum [1]{%
 \ifnum #1\expandafter \@firstoftwo
 \else \expandafter \@secondoftwo
 \fi
}%
\providecommand \@ifx [1]{%
 \ifx #1\expandafter \@firstoftwo
 \else \expandafter \@secondoftwo
 \fi
}%
\providecommand \natexlab [1]{#1}%
\providecommand \enquote  [1]{``#1''}%
\providecommand \bibnamefont  [1]{#1}%
\providecommand \bibfnamefont [1]{#1}%
\providecommand \citenamefont [1]{#1}%
\providecommand \href@noop [0]{\@secondoftwo}%
\providecommand \href [0]{\begingroup \@sanitize@url \@href}%
\providecommand \@href[1]{\@@startlink{#1}\@@href}%
\providecommand \@@href[1]{\endgroup#1\@@endlink}%
\providecommand \@sanitize@url [0]{\catcode `\\12\catcode `\$12\catcode
  `\&12\catcode `\#12\catcode `\^12\catcode `\_12\catcode `\%12\relax}%
\providecommand \@@startlink[1]{}%
\providecommand \@@endlink[0]{}%
\providecommand \url  [0]{\begingroup\@sanitize@url \@url }%
\providecommand \@url [1]{\endgroup\@href {#1}{\urlprefix }}%
\providecommand \urlprefix  [0]{URL }%
\providecommand \Eprint [0]{\href }%
\providecommand \doibase [0]{http://dx.doi.org/}%
\providecommand \selectlanguage [0]{\@gobble}%
\providecommand \bibinfo  [0]{\@secondoftwo}%
\providecommand \bibfield  [0]{\@secondoftwo}%
\providecommand \translation [1]{[#1]}%
\providecommand \BibitemOpen [0]{}%
\providecommand \bibitemStop [0]{}%
\providecommand \bibitemNoStop [0]{.\EOS\space}%
\providecommand \EOS [0]{\spacefactor3000\relax}%
\providecommand \BibitemShut  [1]{\csname bibitem#1\endcsname}%
\let\auto@bib@innerbib\@empty
\bibitem [{\citenamefont {Ye}\ \emph {et~al.}(2018{\natexlab{a}})\citenamefont
  {Ye}, \citenamefont {Kang}, \citenamefont {Liu}, \citenamefont {{von Cube}},
  \citenamefont {Wicker}, \citenamefont {Suzuki}, \citenamefont {Jozwiak},
  \citenamefont {Bostwick}, \citenamefont {Rotenberg}, \citenamefont {Bell},
  \citenamefont {Fu}, \citenamefont {Comin},\ and\ \citenamefont
  {Checkelsky}}]{Ye2018a}%
  \BibitemOpen
  \bibfield  {author} {\bibinfo {author} {\bibfnamefont {L.}~\bibnamefont
  {Ye}}, \bibinfo {author} {\bibfnamefont {M.}~\bibnamefont {Kang}}, \bibinfo
  {author} {\bibfnamefont {J.}~\bibnamefont {Liu}}, \bibinfo {author}
  {\bibfnamefont {F.}~\bibnamefont {{von Cube}}}, \bibinfo {author}
  {\bibfnamefont {C.~R.}\ \bibnamefont {Wicker}}, \bibinfo {author}
  {\bibfnamefont {T.}~\bibnamefont {Suzuki}}, \bibinfo {author} {\bibfnamefont
  {C.}~\bibnamefont {Jozwiak}}, \bibinfo {author} {\bibfnamefont
  {A.}~\bibnamefont {Bostwick}}, \bibinfo {author} {\bibfnamefont
  {E.}~\bibnamefont {Rotenberg}}, \bibinfo {author} {\bibfnamefont {D.~C.}\
  \bibnamefont {Bell}}, \bibinfo {author} {\bibfnamefont {L.}~\bibnamefont
  {Fu}}, \bibinfo {author} {\bibfnamefont {R.}~\bibnamefont {Comin}}, \ and\
  \bibinfo {author} {\bibfnamefont {J.~G.}\ \bibnamefont {Checkelsky}},\
  }\href@noop {} {\bibfield  {journal} {\bibinfo  {journal} {Nature}\ }\textbf
  {\bibinfo {volume} {555}},\ \bibinfo {pages} {638} (\bibinfo {year}
  {2018}{\natexlab{a}})}\BibitemShut {NoStop}%
\bibitem [{\citenamefont {Ye}\ \emph {et~al.}(2018{\natexlab{b}})\citenamefont
  {Ye}, \citenamefont {Chan}, \citenamefont {McDonald}, \citenamefont {Graf},
  \citenamefont {Kang}, \citenamefont {Liu}, \citenamefont {Suzuki},
  \citenamefont {Comin}, \citenamefont {Fu},\ and\ \citenamefont
  {Checkelsky}}]{Ye2018}%
  \BibitemOpen
  \bibfield  {author} {\bibinfo {author} {\bibfnamefont {L.}~\bibnamefont
  {Ye}}, \bibinfo {author} {\bibfnamefont {M.~K.}\ \bibnamefont {Chan}},
  \bibinfo {author} {\bibfnamefont {R.~D.}\ \bibnamefont {McDonald}}, \bibinfo
  {author} {\bibfnamefont {D.}~\bibnamefont {Graf}}, \bibinfo {author}
  {\bibfnamefont {M.}~\bibnamefont {Kang}}, \bibinfo {author} {\bibfnamefont
  {J.}~\bibnamefont {Liu}}, \bibinfo {author} {\bibfnamefont {T.}~\bibnamefont
  {Suzuki}}, \bibinfo {author} {\bibfnamefont {R.}~\bibnamefont {Comin}},
  \bibinfo {author} {\bibfnamefont {L.}~\bibnamefont {Fu}}, \ and\ \bibinfo
  {author} {\bibfnamefont {J.~G.}\ \bibnamefont {Checkelsky}},\ }\href@noop {}
  {\bibfield  {journal} {\bibinfo  {journal} {arXiv:1809.11159}\ } (\bibinfo
  {year} {2018}{\natexlab{b}})}\BibitemShut {NoStop}%
\bibitem [{\citenamefont {Yin}\ \emph {et~al.}(2018)\citenamefont {Yin},
  \citenamefont {Zhang}, \citenamefont {Li}, \citenamefont {Jiang},
  \citenamefont {Chang}, \citenamefont {Zhang}, \citenamefont {Lian},
  \citenamefont {Xiang}, \citenamefont {Belopolski}, \citenamefont {Zheng},
  \citenamefont {Cochran}, \citenamefont {Xu}, \citenamefont {Bian},
  \citenamefont {Liu}, \citenamefont {Chang}, \citenamefont {Lin},
  \citenamefont {Lu}, \citenamefont {Wang}, \citenamefont {Jia}, \citenamefont
  {Wang},\ and\ \citenamefont {Hasan}}]{Yin2018}%
  \BibitemOpen
  \bibfield  {author} {\bibinfo {author} {\bibfnamefont {J.-X.}\ \bibnamefont
  {Yin}}, \bibinfo {author} {\bibfnamefont {S.~S.}\ \bibnamefont {Zhang}},
  \bibinfo {author} {\bibfnamefont {H.}~\bibnamefont {Li}}, \bibinfo {author}
  {\bibfnamefont {K.}~\bibnamefont {Jiang}}, \bibinfo {author} {\bibfnamefont
  {G.}~\bibnamefont {Chang}}, \bibinfo {author} {\bibfnamefont
  {B.}~\bibnamefont {Zhang}}, \bibinfo {author} {\bibfnamefont
  {B.}~\bibnamefont {Lian}}, \bibinfo {author} {\bibfnamefont {C.}~\bibnamefont
  {Xiang}}, \bibinfo {author} {\bibfnamefont {I.}~\bibnamefont {Belopolski}},
  \bibinfo {author} {\bibfnamefont {H.}~\bibnamefont {Zheng}}, \bibinfo
  {author} {\bibfnamefont {T.~A.}\ \bibnamefont {Cochran}}, \bibinfo {author}
  {\bibfnamefont {S.-Y.}\ \bibnamefont {Xu}}, \bibinfo {author} {\bibfnamefont
  {G.}~\bibnamefont {Bian}}, \bibinfo {author} {\bibfnamefont {K.}~\bibnamefont
  {Liu}}, \bibinfo {author} {\bibfnamefont {T.-R.}\ \bibnamefont {Chang}},
  \bibinfo {author} {\bibfnamefont {H.}~\bibnamefont {Lin}}, \bibinfo {author}
  {\bibfnamefont {Z.-Y.}\ \bibnamefont {Lu}}, \bibinfo {author} {\bibfnamefont
  {Z.}~\bibnamefont {Wang}}, \bibinfo {author} {\bibfnamefont {S.}~\bibnamefont
  {Jia}}, \bibinfo {author} {\bibfnamefont {W.}~\bibnamefont {Wang}}, \ and\
  \bibinfo {author} {\bibfnamefont {M.~Z.}\ \bibnamefont {Hasan}},\ }\href@noop
  {} {\bibfield  {journal} {\bibinfo  {journal} {Nature}\ }\textbf {\bibinfo
  {volume} {562}},\ \bibinfo {pages} {91} (\bibinfo {year} {2018})}\BibitemShut
  {NoStop}%
\bibitem [{\citenamefont {Yin}\ \emph {et~al.}(2019)\citenamefont {Yin},
  \citenamefont {Zhang}, \citenamefont {Chang}, \citenamefont {Wang},
  \citenamefont {Tsirkin}, \citenamefont {Guguchia}, \citenamefont {Lian},
  \citenamefont {Zhou}, \citenamefont {Jiang}, \citenamefont {Belopolski},
  \citenamefont {Shumiya}, \citenamefont {Multer}, \citenamefont {Litskevich},
  \citenamefont {Cochran}, \citenamefont {Lin}, \citenamefont {Wang},
  \citenamefont {Neupert}, \citenamefont {Jia}, \citenamefont {Lei},\ and\
  \citenamefont {Hasan}}]{Yin2019}%
  \BibitemOpen
  \bibfield  {author} {\bibinfo {author} {\bibfnamefont {J.-X.}\ \bibnamefont
  {Yin}}, \bibinfo {author} {\bibfnamefont {S.~S.}\ \bibnamefont {Zhang}},
  \bibinfo {author} {\bibfnamefont {G.}~\bibnamefont {Chang}}, \bibinfo
  {author} {\bibfnamefont {Q.}~\bibnamefont {Wang}}, \bibinfo {author}
  {\bibfnamefont {S.~S.}\ \bibnamefont {Tsirkin}}, \bibinfo {author}
  {\bibfnamefont {Z.}~\bibnamefont {Guguchia}}, \bibinfo {author}
  {\bibfnamefont {B.}~\bibnamefont {Lian}}, \bibinfo {author} {\bibfnamefont
  {H.}~\bibnamefont {Zhou}}, \bibinfo {author} {\bibfnamefont {K.}~\bibnamefont
  {Jiang}}, \bibinfo {author} {\bibfnamefont {I.}~\bibnamefont {Belopolski}},
  \bibinfo {author} {\bibfnamefont {N.}~\bibnamefont {Shumiya}}, \bibinfo
  {author} {\bibfnamefont {D.}~\bibnamefont {Multer}}, \bibinfo {author}
  {\bibfnamefont {M.}~\bibnamefont {Litskevich}}, \bibinfo {author}
  {\bibfnamefont {T.~A.}\ \bibnamefont {Cochran}}, \bibinfo {author}
  {\bibfnamefont {H.}~\bibnamefont {Lin}}, \bibinfo {author} {\bibfnamefont
  {Z.}~\bibnamefont {Wang}}, \bibinfo {author} {\bibfnamefont {T.}~\bibnamefont
  {Neupert}}, \bibinfo {author} {\bibfnamefont {S.}~\bibnamefont {Jia}},
  \bibinfo {author} {\bibfnamefont {H.}~\bibnamefont {Lei}}, \ and\ \bibinfo
  {author} {\bibfnamefont {M.~Z.}\ \bibnamefont {Hasan}},\ }\href@noop {}
  {\bibfield  {journal} {\bibinfo  {journal} {Nat. Phys.}\ }\textbf {\bibinfo
  {volume} {15}},\ \bibinfo {pages} {443} (\bibinfo {year} {2019})}\BibitemShut
  {NoStop}%
\bibitem [{\citenamefont {Tang}\ \emph {et~al.}(2011)\citenamefont {Tang},
  \citenamefont {Mei},\ and\ \citenamefont {Wen}}]{Tang2011}%
  \BibitemOpen
  \bibfield  {author} {\bibinfo {author} {\bibfnamefont {E.}~\bibnamefont
  {Tang}}, \bibinfo {author} {\bibfnamefont {J.~W.}\ \bibnamefont {Mei}}, \
  and\ \bibinfo {author} {\bibfnamefont {X.~G.}\ \bibnamefont {Wen}},\
  }\href@noop {} {\bibfield  {journal} {\bibinfo  {journal} {Phys. Rev. Lett.}\
  }\textbf {\bibinfo {volume} {106}},\ \bibinfo {pages} {236802} (\bibinfo
  {year} {2011})}\BibitemShut {NoStop}%
\bibitem [{\citenamefont {Mazin}\ \emph {et~al.}(2014)\citenamefont {Mazin},
  \citenamefont {Jeschke}, \citenamefont {Lechermann}, \citenamefont {Lee},
  \citenamefont {Fink}, \citenamefont {Thomale},\ and\ \citenamefont
  {Valenti}}]{Mazin2014}%
  \BibitemOpen
  \bibfield  {author} {\bibinfo {author} {\bibfnamefont {I.~I.}\ \bibnamefont
  {Mazin}}, \bibinfo {author} {\bibfnamefont {H.~O.}\ \bibnamefont {Jeschke}},
  \bibinfo {author} {\bibfnamefont {F.}~\bibnamefont {Lechermann}}, \bibinfo
  {author} {\bibfnamefont {H.}~\bibnamefont {Lee}}, \bibinfo {author}
  {\bibfnamefont {M.}~\bibnamefont {Fink}}, \bibinfo {author} {\bibfnamefont
  {R.}~\bibnamefont {Thomale}}, \ and\ \bibinfo {author} {\bibfnamefont
  {R.}~\bibnamefont {Valenti}},\ }\href@noop {} {\bibfield  {journal} {\bibinfo
   {journal} {Nat. Commun.}\ }\textbf {\bibinfo {volume} {5}},\ \bibinfo
  {pages} {4261} (\bibinfo {year} {2014})}\BibitemShut {NoStop}%
\bibitem [{\citenamefont {Connell}\ \emph {et~al.}(2012)\citenamefont
  {Connell}, \citenamefont {Isaac}, \citenamefont {Ekanayake}, \citenamefont
  {Strachan},\ and\ \citenamefont {Seo}}]{Connell2012}%
  \BibitemOpen
  \bibfield  {author} {\bibinfo {author} {\bibfnamefont {J.~G.}\ \bibnamefont
  {Connell}}, \bibinfo {author} {\bibfnamefont {B.~J.}\ \bibnamefont {Isaac}},
  \bibinfo {author} {\bibfnamefont {G.~B.}\ \bibnamefont {Ekanayake}}, \bibinfo
  {author} {\bibfnamefont {D.~R.}\ \bibnamefont {Strachan}}, \ and\ \bibinfo
  {author} {\bibfnamefont {S.~S.~A.}\ \bibnamefont {Seo}},\ }\href@noop {}
  {\bibfield  {journal} {\bibinfo  {journal} {Appl. Phys. Lett.}\ }\textbf
  {\bibinfo {volume} {101}},\ \bibinfo {pages} {251607} (\bibinfo {year}
  {2012})}\BibitemShut {NoStop}%
\bibitem [{\citenamefont {Hallsteinsen}\ \emph {et~al.}(2016)\citenamefont
  {Hallsteinsen}, \citenamefont {Nord}, \citenamefont {Bolstad}, \citenamefont
  {Vullum}, \citenamefont {Boschker}, \citenamefont {Longo}, \citenamefont
  {Takahashi}, \citenamefont {Holmestad}, \citenamefont {Lippmaa},\ and\
  \citenamefont {Tybell}}]{Hallsteinsen2016}%
  \BibitemOpen
  \bibfield  {author} {\bibinfo {author} {\bibfnamefont {I.}~\bibnamefont
  {Hallsteinsen}}, \bibinfo {author} {\bibfnamefont {M.}~\bibnamefont {Nord}},
  \bibinfo {author} {\bibfnamefont {T.}~\bibnamefont {Bolstad}}, \bibinfo
  {author} {\bibfnamefont {P.-E.}\ \bibnamefont {Vullum}}, \bibinfo {author}
  {\bibfnamefont {J.~E.}\ \bibnamefont {Boschker}}, \bibinfo {author}
  {\bibfnamefont {P.}~\bibnamefont {Longo}}, \bibinfo {author} {\bibfnamefont
  {R.}~\bibnamefont {Takahashi}}, \bibinfo {author} {\bibfnamefont
  {R.}~\bibnamefont {Holmestad}}, \bibinfo {author} {\bibfnamefont
  {M.}~\bibnamefont {Lippmaa}}, \ and\ \bibinfo {author} {\bibfnamefont
  {T.}~\bibnamefont {Tybell}},\ }\href@noop {} {\bibfield  {journal} {\bibinfo
  {journal} {Cryst. Growth Des.}\ }\textbf {\bibinfo {volume} {16}},\ \bibinfo
  {pages} {2357} (\bibinfo {year} {2016})}\BibitemShut {NoStop}%
\bibitem [{\citenamefont {Woo}\ \emph {et~al.}(2015)\citenamefont {Woo},
  \citenamefont {Jeong}, \citenamefont {Lee}, \citenamefont {Seo},
  \citenamefont {Lacotte}, \citenamefont {David}, \citenamefont {Kim},
  \citenamefont {Prellier}, \citenamefont {Kim},\ and\ \citenamefont
  {Choi}}]{Woo2015}%
  \BibitemOpen
  \bibfield  {author} {\bibinfo {author} {\bibfnamefont {S.}~\bibnamefont
  {Woo}}, \bibinfo {author} {\bibfnamefont {H.}~\bibnamefont {Jeong}}, \bibinfo
  {author} {\bibfnamefont {S.~A.}\ \bibnamefont {Lee}}, \bibinfo {author}
  {\bibfnamefont {H.}~\bibnamefont {Seo}}, \bibinfo {author} {\bibfnamefont
  {M.}~\bibnamefont {Lacotte}}, \bibinfo {author} {\bibfnamefont
  {A.}~\bibnamefont {David}}, \bibinfo {author} {\bibfnamefont {H.~Y.}\
  \bibnamefont {Kim}}, \bibinfo {author} {\bibfnamefont {W.}~\bibnamefont
  {Prellier}}, \bibinfo {author} {\bibfnamefont {Y.}~\bibnamefont {Kim}}, \
  and\ \bibinfo {author} {\bibfnamefont {W.~S.}\ \bibnamefont {Choi}},\
  }\href@noop {} {\bibfield  {journal} {\bibinfo  {journal} {Sci. Rep.}\
  }\textbf {\bibinfo {volume} {5}},\ \bibinfo {pages} {8822} (\bibinfo {year}
  {2015})}\BibitemShut {NoStop}%
\bibitem [{\citenamefont {Stenstr\"om}(1972)}]{Stenstrom1972}%
  \BibitemOpen
  \bibfield  {author} {\bibinfo {author} {\bibfnamefont {B.}~\bibnamefont
  {Stenstr\"om}},\ }\href@noop {} {\bibfield  {journal} {\bibinfo  {journal}
  {Phys. Scripta}\ }\textbf {\bibinfo {volume} {6}},\ \bibinfo {pages} {214}
  (\bibinfo {year} {1972})}\BibitemShut {NoStop}%
\bibitem [{\citenamefont {Usami}(1978)}]{Usami1978}%
  \BibitemOpen
  \bibfield  {author} {\bibinfo {author} {\bibfnamefont {K.}~\bibnamefont
  {Usami}},\ }\href@noop {} {\bibfield  {journal} {\bibinfo  {journal} {J.
  Phys. Soc. Jpn.}\ }\textbf {\bibinfo {volume} {45}},\ \bibinfo {pages} {466}
  (\bibinfo {year} {1978})}\BibitemShut {NoStop}%
\bibitem [{\citenamefont {Hartmann}\ and\ \citenamefont
  {W{\"{a}}ppling}(1987)}]{Hartmann1987}%
  \BibitemOpen
  \bibfield  {author} {\bibinfo {author} {\bibfnamefont {O.}~\bibnamefont
  {Hartmann}}\ and\ \bibinfo {author} {\bibfnamefont {R.}~\bibnamefont
  {W{\"{a}}ppling}},\ }\href@noop {} {\bibfield  {journal} {\bibinfo  {journal}
  {Phys. Scripta}\ }\textbf {\bibinfo {volume} {35}},\ \bibinfo {pages} {499}
  (\bibinfo {year} {1987})}\BibitemShut {NoStop}%
\bibitem [{\citenamefont {Yamamoto}(1966)}]{Yamamoto1966}%
  \BibitemOpen
  \bibfield  {author} {\bibinfo {author} {\bibfnamefont {H.}~\bibnamefont
  {Yamamoto}},\ }\href@noop {} {\bibfield  {journal} {\bibinfo  {journal} {J.
  Phys. Soc. Jpn.}\ }\textbf {\bibinfo {volume} {21}},\ \bibinfo {pages} {1058}
  (\bibinfo {year} {1966})}\BibitemShut {NoStop}%
\bibitem [{\citenamefont {Dj\'ega-Mariadassou}\ \emph
  {et~al.}(1966)\citenamefont {Dj\'ega-Mariadassou}, \citenamefont {Lecocq},
  \citenamefont {Trumpy}, \citenamefont {Tr\"aff},\ and\ \citenamefont
  {{\O}stergaard}}]{Djega-Mariadassou1966}%
  \BibitemOpen
  \bibfield  {author} {\bibinfo {author} {\bibfnamefont {C.}~\bibnamefont
  {Dj\'ega-Mariadassou}}, \bibinfo {author} {\bibfnamefont {P.}~\bibnamefont
  {Lecocq}}, \bibinfo {author} {\bibfnamefont {G.}~\bibnamefont {Trumpy}},
  \bibinfo {author} {\bibfnamefont {J.}~\bibnamefont {Tr\"aff}}, \ and\
  \bibinfo {author} {\bibfnamefont {P.}~\bibnamefont {{\O}stergaard}},\
  }\href@noop {} {\bibfield  {journal} {\bibinfo  {journal} {Nuovo Cimento B}\
  }\textbf {\bibinfo {volume} {46}},\ \bibinfo {pages} {35} (\bibinfo {year}
  {1966})}\BibitemShut {NoStop}%
\bibitem [{\citenamefont {Yamaguchi}\ and\ \citenamefont
  {Watanabe}(1967)}]{Yamaguchi1967}%
  \BibitemOpen
  \bibfield  {author} {\bibinfo {author} {\bibfnamefont {K.}~\bibnamefont
  {Yamaguchi}}\ and\ \bibinfo {author} {\bibfnamefont {H.}~\bibnamefont
  {Watanabe}},\ }\href@noop {} {\bibfield  {journal} {\bibinfo  {journal} {J.
  Phys. Soc. Jpn.}\ }\textbf {\bibinfo {volume} {22}},\ \bibinfo {pages} {1210}
  (\bibinfo {year} {1967})}\BibitemShut {NoStop}%
\bibitem [{\citenamefont {Kulshreshtha}\ and\ \citenamefont
  {Raj}(1981)}]{Kulshreshtha1981}%
  \BibitemOpen
  \bibfield  {author} {\bibinfo {author} {\bibfnamefont {S.~K.}\ \bibnamefont
  {Kulshreshtha}}\ and\ \bibinfo {author} {\bibfnamefont {P.}~\bibnamefont
  {Raj}},\ }\href@noop {} {\bibfield  {journal} {\bibinfo  {journal} {J. Phys.
  F. Met. Phys.}\ }\textbf {\bibinfo {volume} {11}},\ \bibinfo {pages} {281}
  (\bibinfo {year} {1981})}\BibitemShut {NoStop}%
\bibitem [{\citenamefont {Haggstrom}\ \emph {et~al.}(1975)\citenamefont
  {Haggstrom}, \citenamefont {Ericsson}, \citenamefont {W{\"{a}}ppling},\ and\
  \citenamefont {Chandra}}]{Haggstrom1975}%
  \BibitemOpen
  \bibfield  {author} {\bibinfo {author} {\bibfnamefont {L.}~\bibnamefont
  {Haggstrom}}, \bibinfo {author} {\bibfnamefont {T.}~\bibnamefont {Ericsson}},
  \bibinfo {author} {\bibfnamefont {R.}~\bibnamefont {W{\"{a}}ppling}}, \ and\
  \bibinfo {author} {\bibfnamefont {K.}~\bibnamefont {Chandra}},\ }\href@noop
  {} {\bibfield  {journal} {\bibinfo  {journal} {Phys. Scripta}\ }\textbf
  {\bibinfo {volume} {11}},\ \bibinfo {pages} {47} (\bibinfo {year}
  {1975})}\BibitemShut {NoStop}%
\bibitem [{\citenamefont {Wang}\ \emph {et~al.}(2005)\citenamefont {Wang},
  \citenamefont {Li}, \citenamefont {Naughton}, \citenamefont {Gu},
  \citenamefont {Uchida},\ and\ \citenamefont {Ong}}]{Wang2005}%
  \BibitemOpen
  \bibfield  {author} {\bibinfo {author} {\bibfnamefont {Y.}~\bibnamefont
  {Wang}}, \bibinfo {author} {\bibfnamefont {L.}~\bibnamefont {Li}}, \bibinfo
  {author} {\bibfnamefont {M.}~\bibnamefont {Naughton}}, \bibinfo {author}
  {\bibfnamefont {G.}~\bibnamefont {Gu}}, \bibinfo {author} {\bibfnamefont
  {S.}~\bibnamefont {Uchida}}, \ and\ \bibinfo {author} {\bibfnamefont
  {N.}~\bibnamefont {Ong}},\ }\href@noop {} {\bibfield  {journal} {\bibinfo
  {journal} {Phys. Rev. Lett.}\ }\textbf {\bibinfo {volume} {95}},\ \bibinfo
  {pages} {247002} (\bibinfo {year} {2005})}\BibitemShut {NoStop}%
\bibitem [{\citenamefont {Inoue}\ \emph {et~al.}(2019)\citenamefont {Inoue},
  \citenamefont {Han}, \citenamefont {Hu}, \citenamefont {Suzuki},
  \citenamefont {Liu},\ and\ \citenamefont {Checkelsky}}]{Inoue2019}%
  \BibitemOpen
  \bibfield  {author} {\bibinfo {author} {\bibfnamefont {H.}~\bibnamefont
  {Inoue}}, \bibinfo {author} {\bibfnamefont {M.}~\bibnamefont {Han}}, \bibinfo
  {author} {\bibfnamefont {M.}~\bibnamefont {Hu}}, \bibinfo {author}
  {\bibfnamefont {T.}~\bibnamefont {Suzuki}}, \bibinfo {author} {\bibfnamefont
  {J.}~\bibnamefont {Liu}}, \ and\ \bibinfo {author} {\bibfnamefont {J.~G.}\
  \bibnamefont {Checkelsky}},\ }\href@noop {} {\bibfield  {journal} {\bibinfo
  {journal} {arXiv:1904.02582}\ } (\bibinfo {year} {2019})}\BibitemShut
  {NoStop}%
\bibitem [{\citenamefont {{\v{S}}mejkal}\ \emph {et~al.}(2017)\citenamefont
  {{\v{S}}mejkal}, \citenamefont {Jungwirth},\ and\ \citenamefont
  {Sinova}}]{Smejkal2017}%
  \BibitemOpen
  \bibfield  {author} {\bibinfo {author} {\bibfnamefont {L.}~\bibnamefont
  {{\v{S}}mejkal}}, \bibinfo {author} {\bibfnamefont {T.}~\bibnamefont
  {Jungwirth}}, \ and\ \bibinfo {author} {\bibfnamefont {J.}~\bibnamefont
  {Sinova}},\ }\href@noop {} {\bibfield  {journal} {\bibinfo  {journal} {Phys.
  Status Solidi-R}\ }\textbf {\bibinfo {volume} {11}},\ \bibinfo {pages}
  {1700044} (\bibinfo {year} {2017})}\BibitemShut {NoStop}%
\bibitem [{\citenamefont {{\v{S}}mejkal}\ \emph {et~al.}(2018)\citenamefont
  {{\v{S}}mejkal}, \citenamefont {Mokrousov}, \citenamefont {Yan},\ and\
  \citenamefont {MacDonald}}]{Smejkal2018}%
  \BibitemOpen
  \bibfield  {author} {\bibinfo {author} {\bibfnamefont {L.}~\bibnamefont
  {{\v{S}}mejkal}}, \bibinfo {author} {\bibfnamefont {Y.}~\bibnamefont
  {Mokrousov}}, \bibinfo {author} {\bibfnamefont {B.}~\bibnamefont {Yan}}, \
  and\ \bibinfo {author} {\bibfnamefont {A.~H.}\ \bibnamefont {MacDonald}},\
  }\href@noop {} {\bibfield  {journal} {\bibinfo  {journal} {Nat. Phys.}\
  }\textbf {\bibinfo {volume} {14}},\ \bibinfo {pages} {242} (\bibinfo {year}
  {2018})}\BibitemShut {NoStop}%
\end{thebibliography}%


%

\end{document}